\documentclass[a4paper,11pt]{article}
\usepackage{pos}

\title{The throughput calibration of the VERITAS telescopes}

\author*[a,b,c]{M. Nievas Rosillo}
\author{the VERITAS collaboration}

\affiliation[a]{Instituto de Astrof\'isica de Canarias (IAC) \\ C/Via Lactea S/N E-38205 La Laguna, Tenerife, Spain}

\affiliation[b]{Universidad de La Laguna, Dept. Astrof\'isica (ULL) \\
Av. Astrofisico Francisco Sánchez, S/N, E-38206 La Laguna, Tenerife, Spain}

\affiliation[c]{Deutsches Elektronen-Synchrotron (DESY) \\ Platanenallee 6, Zeuthen, Germany, Spain}

\emailAdd{mnievas@iac.es}

\abstract{Imaging atmospheric Cherenkov telescopes are continuously exposed to varying weather conditions that have short and long-term effects on their response to Cherenkov light from extensive air showers. This work presents the implementation of a throughput calibration method for the VERITAS telescopes taking into account changes in the optical response and detector performance over time. Different methods to measure the total throughput of the instrument, which depend on mirror reflectivites and PMT camera gain and efficiency, are discussed as well as the effect of its evolution on energy thresholds, effective collection areas, and energy reconstruction. The application of this calibration in the VERITAS data analysis chain is discussed, including the validation using Monte Carlo simulations and observations of the Crab Nebula.}

\FullConference{37$^{\rm{th}}$ International Cosmic Ray Conference (ICRC 2021)\\
		July 12th -- 23rd, 2021\\
		Online -- Berlin, Germany}

\begin{document}
\maketitle

\section{Main Objetives}

The Very Energetic Radiation Imaging Telescope  Array  System  (VERITAS) is a ground-based very-high-energy (VHE, $E>100\,\mathrm{GeV}$) instrument located at the Fred Lawrence Whipple Observatory (FLWO) in southern Arizona (31 40N, 110 57W,  1.3km a.s.l.). It consists of four $12\,\mathrm{m}$ Imaging Air Cherenkov Telescopes (IACT) with a Davies-Cotton design. The array was commissioned in 2007 \cite{2007JPhCS..60...34K} and upgraded in 2012 \cite{2011ICRC....9...14K}.

%The study of VHE $\gamma$-rays with IACTs relies on detailed Monte Carlo (MC) simulations of the production of Extensive Air Showers (EAS) in the atmosphere and the detection of its Cherenkov flashes with the telescopes. 

%The response of Cherenkov telescopes to EAS evolves over time.
IACTs are exposed to varying weather conditions that degrade the mirrors. Additionally, cameras are usually composed of photomultiplier tubes (PMT), which work at high voltages and degrade as charge accumulates. This work discusses the monitoring of the VERITAS throughput or efficiency converting incoming Cherenkov light into measured signals, the correction of Monte Carlo (MC) simulations to changes in this parameter and the generation of the corresponding throughput-calibrated Instrument Response Functions (IRF).

\section{Throughput measurements and calibration}

The total telescope throughput correlates with the efficiency of camera converting light into measurable signals and primary mirror reflectivity:

\begin{itemize}
\item\ \textbf{\em PMT gains and quantum efficiency (QE)} affect the conversion of photons into measurable signals in the camera. VERITAS simulations use an average value of the PMT gain $G$ for each telescope ($G_{MC}$). %Relative gain differences between PMTs are measured daily using uniformly illuminated camera events (`LED Flasher runs') and corrected during the calibration of real Cherenkov data.
%Flasher runs also provide an estimation of the average camera gain evolution over time. Based on this, relative `gain factors' $g=\langle G \rangle>/G_{MC}$ can be produced. They are required to correct the MC for gain evolution over time. 
Uniformly illuminated camera events (`LED Flasher runs' \cite{2010NIMPA.612..278H}) are collected each night to be used during data calibration. They provide an estimation of the relative gains of each pixel and statistical analysis provides also an estimation of the camera gains $G$, which are used to compute a correction term $g=G/G_{MC}$.
%At the beginning of the observing season, HV settings of the cameras are adjusted taking into account these measurements. This reduces long term gain drifts to $\lesssim 10\%$. %For these camera events, the mean pulse charge at each PMT is statistically proportional to the mean number of phoelectrons at the first dynode.
\item\ \textbf{\em Reflectivity} is measured using a wide-field CCD camera attached to the primary mirror and a spectralon target plate over the focal plane \cite{2021APh...12802556A}. Using CCD images of bright stars and their reflection on the spectralon, we estimate the reflectivity of the primary mirror $T$. The ratio to the wavelength-averaged reflectivities assumed in the MC $T_{MC}$ yields a correction term $t=T/T_{MC}$. % The comparison of the primary mirror reflectivity to what is simulated in the MC model is used to produce relative `optical throughput factors' $t=T/T_{MC}$ that are needed to correct the simulations for reflectivity changes in the primary mirror.
\end{itemize}

The \textbf{\em total throughput} correction ($s$-factor), combination of camera and primary mirror factors, is then estimated as $s = g \times t$. 

\begin{figure}
\centering
\includegraphics[width=0.9\linewidth]{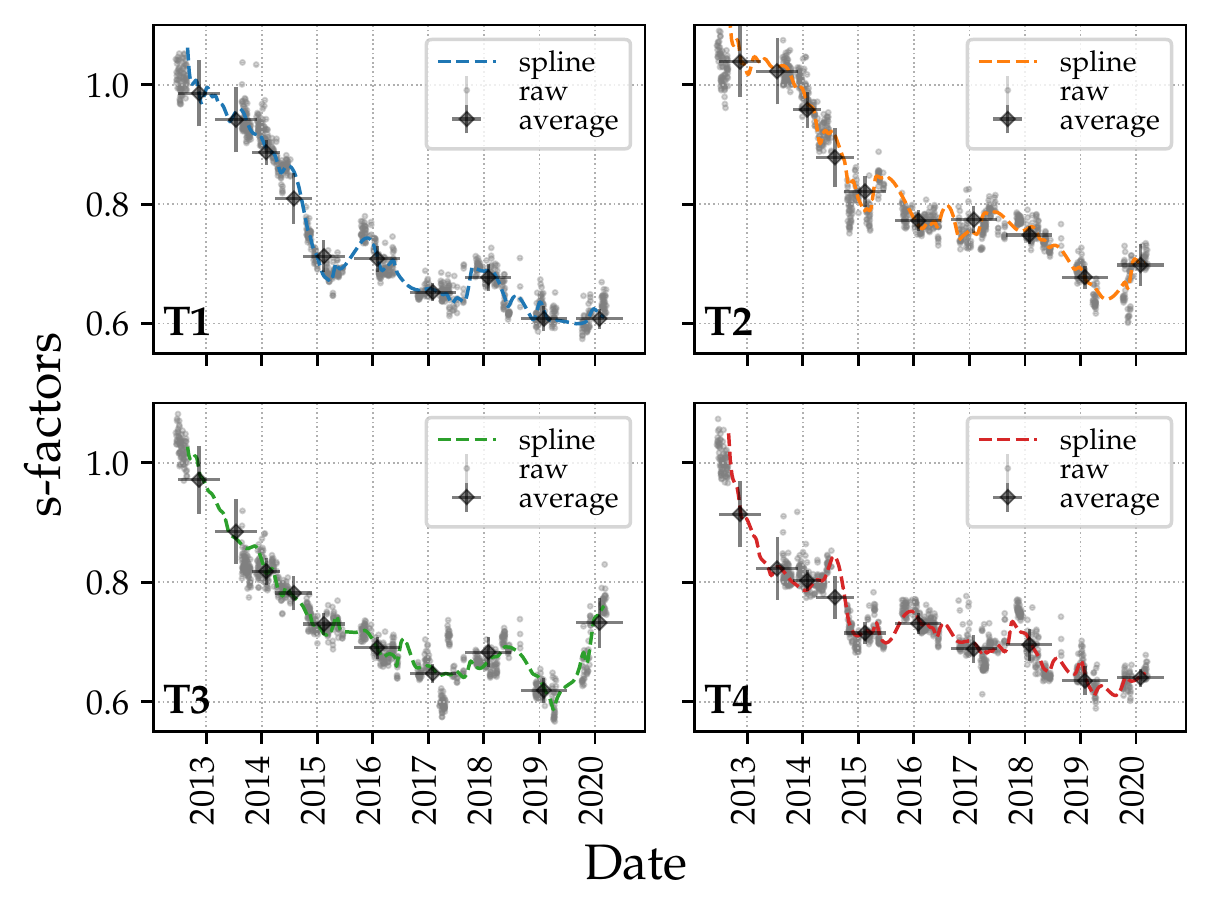}
%\caption{Relative throughput (gain + reflectivity) with respect to the throughput assumed in the Monte Carlo simulations of VERITAS. }
\caption{Evolution of $s$-factors for the four VERITAS telescopes}
\label{fig:total_throughput}
\end{figure}

To make the computation of IRFs feasible, we define instrument periods with maximum oscillations of throughput of $10\%$ for which we can compute an average $s$-factor. Calculated for each telescope and period, these $s$-factors allow us to correct the reconstructed PMT signals of the simulated $\gamma$-ray events, before the PMT traces are integrated and shower images cleaned. Figure \ref{fig:total_throughput} shows the $s$-factors for each telescope. Note they do not decrease monotonically since mirrors are occasionally replaced, which brings the reflectivity up. % \ref{fig:total_throughput}. % shows the resulting $s$-factors.

\section{Analysis packages}

The VERITAS collaboration maintains two analysis packages: Eventdisplay \cite{2017ICRC...35..747M} and VEGAS \cite{2008ICRC....3.1385C}. While the analysis methods and throughput corrections are similar and have been validated with both packagse, we focus on results obtained with Eventdisplay.

\section{Effects of throughput changes on the analysis}

%----------------------------------------------------------------------------------------
%	RESULTS 
%----------------------------------------------------------------------------------------

%\section*{Throughput measurements}

The analysis of $\gamma$-ray showers is impacted by throughput degradation in two ways: 

\begin{enumerate}
\item Energy reconstruction bias. 
\item Loss of events at low energies, near the threshold. 
\end{enumerate}

\subsection{Energy reconstruction}

\begin{figure}
\centering
\includegraphics[width=0.7\linewidth]{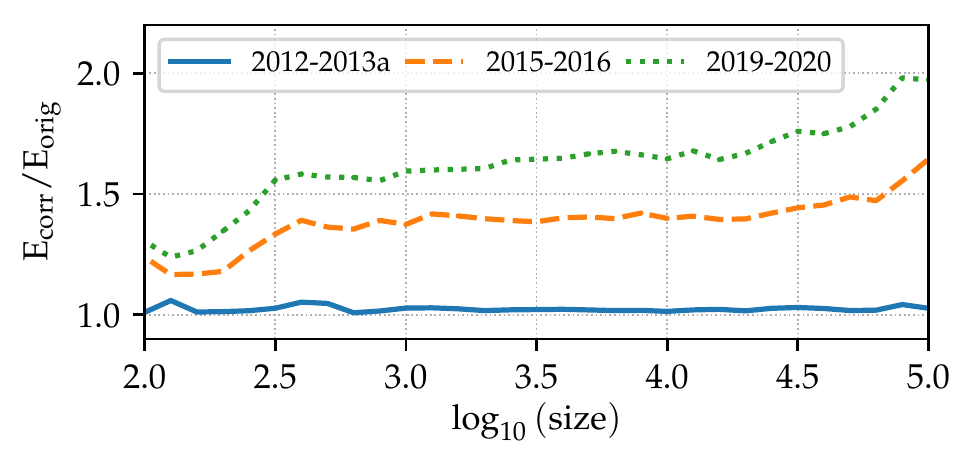}
%\caption{Impact of instrument ageing on the energy lookup tables.}
\caption{Ratio of throughput-calibrated $\mathrm{E_{corr}}$ to uncalibrated $\mathrm{E_{orig}}$ reconstructed energies (for telescope T1) as a function of size of the images, fixing all the other parameters involved in the construction of the tables (noise, shower distance, zenith, etc). Three different IRF periods are shown, showing increasing ratios of throughput-calibrated energy to uncalibrated energy.}
\label{fig:lookup_tables}
%\vspace{1cm}
\end{figure}

The energy reconstruction considered in this work is based on lookup tables, which encode the dependency of the event energy on size (integrated particle shower charge over the camera) among other parameters. These energy lookup tables are filled with throughput-calibrated MC events and then applied to real data. Throughput drops cause events to deposit a smaller charge in the camera pixels, therefore the reconstructed events sizes are smaller the lower the throughput is. The effect of throughput evolution on them is presented in Figure \ref{fig:lookup_tables}. %In the following plot we show the effect of throughput corrections on these tables, fixing all the parameters except the size and computing the ratio of throughput-corrected and uncorrected reconstructed energies. %The effect of throughput-corrections in energy reconstruction is shown in the following figure, where for a specific set of parameters $\mathrm{E_{corr}}$ is the energy reconstructed after throughput-calibration and $\mathrm{E_{orig}}$ the corresponding energy without throughput corrections.

\subsection{Sensitivity and Energy Threshold}

\begin{figure}
\centering
\includegraphics[width=0.54\linewidth]{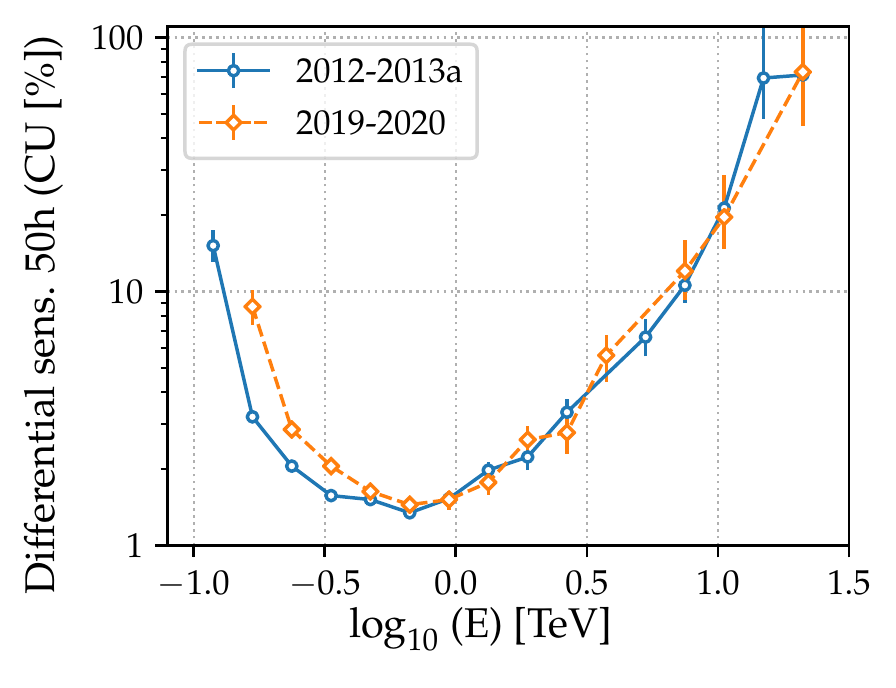} \includegraphics[width=0.44\linewidth]{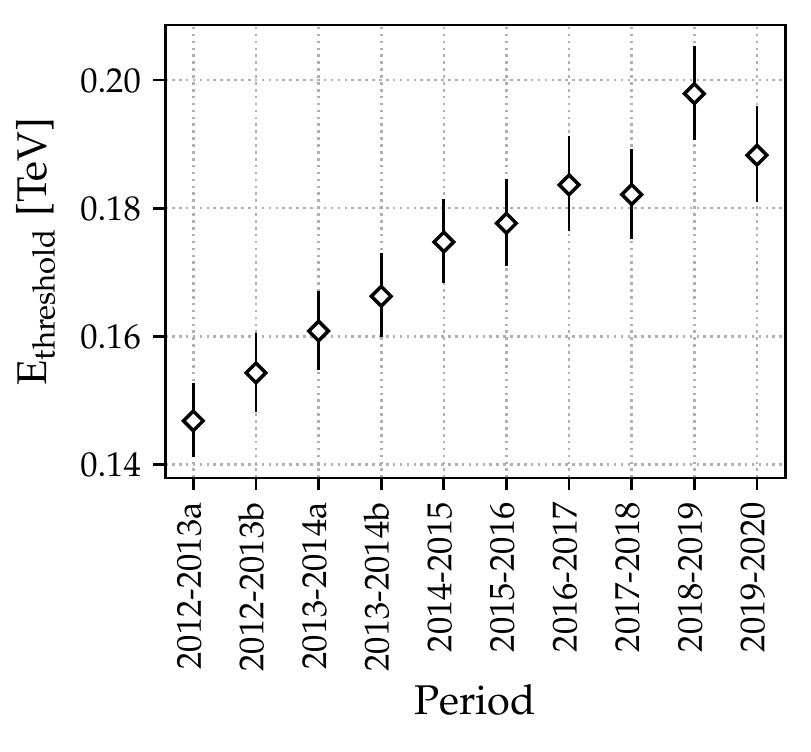}
%\caption{Performance metrics after taking into account instrument ageing. Left: Differential sensitivity (measured for $50\,\mathrm{h}$ in Crab Units for 3 observation campaigns. Right: Evolution of the analysis energy threshold for a fixed set of cuts (which include size cuts).}
\caption{{\bf Left:} {\em Differential sensitivity} during two campaigns, measured in Crab Nebula flux units; {\bf Right:} {\em Energy threshold}. They were obtained using Crab Nebula data and for a typical set of analysis cuts.}
\label{fig:sensitivity_threshold}
\end{figure}

Decreased throughput causes loss of events near the energy threshold as signals are unable to pass trigger and analysis cuts. % In addition, some more are misclassified due to the smaller signal-to-noise ratios. 
The effect is a worsening of the sensitivity, mostly at low energies, and an increase of the analysis energy threshold. \textbf{\em Differential sensitivity} is defined as the minimum flux (in Crab Units) per energy bin which can be detected in $50\,\mathrm{h}$ with a significance of $5\,\sigma$ using Li\&Ma statistics \cite{1983ApJ...272..317L}. \textbf{\em Energy threshold} is the energy at which the effective area becomes $10\%$ of the maximum effective area of the array. The evolution of both parameters is shown in Figure \ref{fig:sensitivity_threshold}.

\section{Reconstruction of the Crab Nebula spectrum}

\begin{figure}%\vspace{1cm}
\centering
\includegraphics[width=0.9\linewidth]{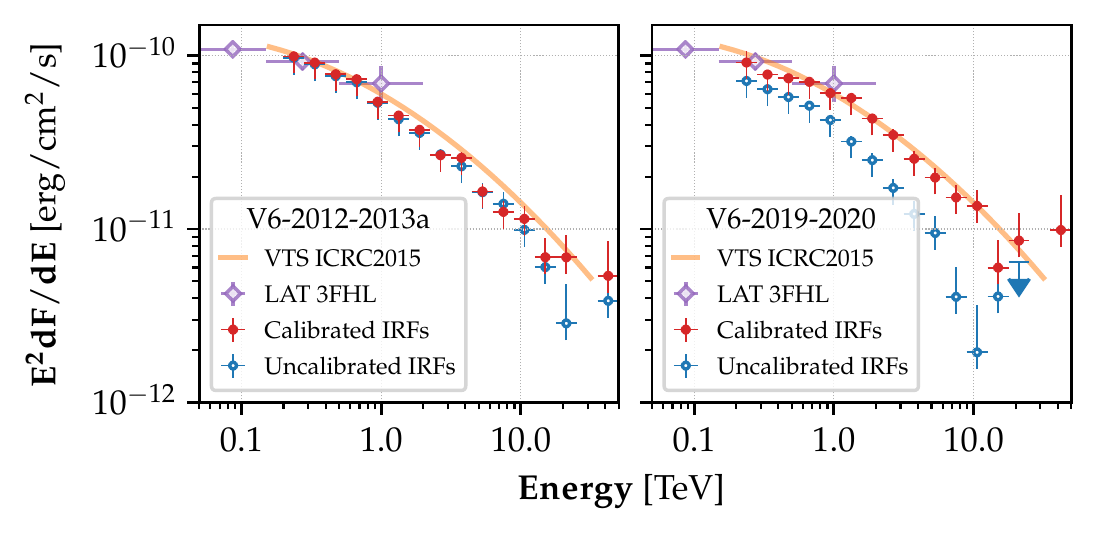}
%\caption{}
\caption{Reconstructed spectra of the Crab Nebula with VERITAS. Blue open points show the spectrum obtained without throughput calibration while red filled points are the reconstructed spectrum when the MCs and IRFs are calibrated. For reference, Fermi-LAT's 4FGL \cite{2020ApJS..247...33A} spectral points for the Crab Nebula are shown as open purple circles and the VERITAS Crab Spectrum presented in \cite{2015ICRC...34..792M} is shown in orange.}
\label{fig:crab_spectra}
\end{figure}

The Crab Nebula is often used as a benchmark source to test the performance of IACTs. It is a bright $\gamma$-ray object, visible from both hemispheres and its spectrum extends throughout the entire VHE band with no significant flux variability having been observed. Figure \ref{fig:crab_spectra} shows the reconstructed spectrum of the Crab Nebula in two campaigns with significantly different telescope throughput. As it is seen from the figure, throughput-calibration of VERITAS MCs and the corresponding IRF allows us to recover nominal fluxes from the Crab Nebula, therefore confirming that the throughput calibration technique described in this document is properly working.

%Throughput variations have an impact on the event energies and reconstructed fluxes from astrophysical sources. % if they are not corrected in the analysis. This affects both the reconstructed spectrum and the light curves. 

%The next plot collects the integral fluxes of the Crab Nebula over time, reconstructed with the VERITAS array. They have been obtained using IRFs that have been produced with throughput-corrected MC simulations. The right panel shows the histogram of flux measurements obtained if no throughput corrections are applied to the IRFs (dashed gray curve) and if they are applied (solid filled black histogram). Fluxes for the individual IRF periods are shown in colors.

%\begin{center}\vspace{1cm}
%\centering
%\includegraphics[width=1\linewidth]{{figures/LC500GeV_Crab}.pdf}
%\caption{}
%\label{fig:effective_area}
%\end{center}

%The reconstructed spectrum of the Crab Nebula, as seen by VERITAS during two campaigns, is presented below. Blue open points show the spectrum obtained without throughput calibration while red filled points the reconstructed spectrum when the MCs and IRFs are calibrated. For reference, Fermi-LAT's 4FGL \cite{2020ApJS..247...33A} spectral points for the Crab Nebula are show as open purple circles and the VERITAS Crab Spectrum presented in \cite{2015ICRC...34..792M} is shown in orange.

%----------------------------------------------------------------------------------------
%	CONCLUSIONS
%----------------------------------------------------------------------------------------

\section{Conclusions}

\begin{itemize}
\item IACT are subject to varying weather conditions which degrade mirrors and PMTs. 
\item Throughput calibration is essential to correctly reconstruct $\gamma$-ray shower parameters.
\item The monitoring of throughput in VERITAS is presented together with corrections applied to simulated events. %This allow us to  reconstruct the correct properties of Cherenkov shower images.
\item The impact of throughput variations in VERITAS performance is evaluated using real observations from the Crab Nebula, a standard candle in VHE $\gamma$-ray astrophysics.
\item The method works well and is now standard in VERITAS.
\end{itemize}

%% Full authors list (ONLY FOR COLLABORATIONS)
%\clearpage
%\section*{Full Authors List: \Coll\ Collaboration}
%
%\noindent \textbf{Note comment afterwards:} Collaborations have the possibility to provide an authors list in xml format which will be used while generating the DOI entries making the full authors list searchable in databases like Inspire HEP. For instructions please go to icrc2021.desy.de/proceedings or contact us under icrc2021proc@desy.de.\\
%
%\scriptsize
%\noindent
%first.author$^1$, 
%second.author$^2$, 
%third.author$^3$ % .... more names
%and 
%last.author$^{n}$ \\
%
%\noindent
%$^1$first.affiliation.
%$^2$second.affiliation. % .... more affiliation
%$^{m}$last.affiliation.

\nocite{*} % Print all references regardless of whether they were cited in the poster or not
%\bibliographystyle{plain} % Plain referencing style
%\bibliography{sample} % Use the example bibliography file sample.bib

%\bibliographystyle{aa} % style aa.bst
\bibliographystyle{unsrt}
\bibliography{sample} % your references Yourfile.bib

%----------------------------------------------------------------------------------------
%	ACKNOWLEDGEMENTS
%----------------------------------------------------------------------------------------

\section*{Acknowledgements}

This research is supported by grants from the U.S. Department of Energy Office of Science, the U.S. National Science Foundation and the Smithsonian Institution, by NSERC in Canada, and by the Helmholtz Association in Germany. This research used resources provided by the Open Science Grid, which is supported by the National Science Foundation and the U.S. Department of Energy's Office of Science, and resources of the National Energy Research Scientific Computing Center (NERSC), a U.S. Department of Energy Office of Science User Facility operated under Contract No. DE-AC02-05CH11231. We acknowledge the excellent work of the technical support staff at the Fred Lawrence Whipple Observatory and at the collaborating institutions in the construction and operation of the instrument.

%----------------------------------------------------------------------------------------
\clearpage
\section*{Full Authors List: VERITAS\ Collaboration}

\scriptsize
\noindent
C.~B.~Adams$^{1}$,
A.~Archer$^{2}$,
W.~Benbow$^{3}$,
A.~Brill$^{1}$,
J.~H.~Buckley$^{4}$,
M.~Capasso$^{5}$,
J.~L.~Christiansen$^{6}$,
A.~J.~Chromey$^{7}$, 
M.~Errando$^{4}$,
A.~Falcone$^{8}$,
K.~A.~Farrell$^{9}$,
Q.~Feng$^{5}$,
G.~M.~Foote$^{10}$,
L.~Fortson$^{11}$,
A.~Furniss$^{12}$,
A.~Gent$^{13}$,
G.~H.~Gillanders$^{14}$,
C.~Giuri$^{15}$,
O.~Gueta$^{15}$,
D.~Hanna$^{16}$,
O.~Hervet$^{17}$,
J.~Holder$^{10}$,
B.~Hona$^{18}$,
T.~B.~Humensky$^{1}$,
W.~Jin$^{19}$,
P.~Kaaret$^{20}$,
M.~Kertzman$^{2}$,
D.~Kieda$^{18}$,
T.~K.~Kleiner$^{15}$,
S.~Kumar$^{16}$,
M.~J.~Lang$^{14}$,
M.~Lundy$^{16}$,
G.~Maier$^{15}$,
C.~E~McGrath$^{9}$,
P.~Moriarty$^{14}$,
R.~Mukherjee$^{5}$,
D.~Nieto$^{21}$,
M.~Nievas-Rosillo$^{15}$,
S.~O'Brien$^{16}$,
R.~A.~Ong$^{22}$,
A.~N.~Otte$^{13}$,
S.~R. Patel$^{15}$,
S.~Patel$^{20}$,
K.~Pfrang$^{15}$,
M.~Pohl$^{23,15}$,
R.~R.~Prado$^{15}$,
E.~Pueschel$^{15}$,
J.~Quinn$^{9}$,
K.~Ragan$^{16}$,
P.~T.~Reynolds$^{24}$,
D.~Ribeiro$^{1}$,
E.~Roache$^{3}$,
J.~L.~Ryan$^{22}$,
I.~Sadeh$^{15}$,
M.~Santander$^{19}$,
G.~H.~Sembroski$^{25}$,
R.~Shang$^{22}$,
D.~Tak$^{15}$,
V.~V.~Vassiliev$^{22}$,
A.~Weinstein$^{7}$,
D.~A.~Williams$^{17}$,
and 
T.~J.~Williamson$^{10}$\\
\noindent
$^{1}${Physics Department, Columbia University, New York, NY 10027, USA}
$^{2}${Department of Physics and Astronomy, DePauw University, Greencastle, IN 46135-0037, USA}
$^{3}${Center for Astrophysics $|$ Harvard \& Smithsonian, Cambridge, MA 02138, USA}
$^{4}${Department of Physics, Washington University, St. Louis, MO 63130, USA}
$^{5}${Department of Physics and Astronomy, Barnard College, Columbia University, NY 10027, USA}
$^{6}${Physics Department, California Polytechnic State University, San Luis Obispo, CA 94307, USA} 
$^{7}${Department of Physics and Astronomy, Iowa State University, Ames, IA 50011, USA}
$^{8}${Department of Astronomy and Astrophysics, 525 Davey Lab, Pennsylvania State University, University Park, PA 16802, USA}
$^{9}${School of Physics, University College Dublin, Belfield, Dublin 4, Ireland}
$^{10}${Department of Physics and Astronomy and the Bartol Research Institute, University of Delaware, Newark, DE 19716, USA}
$^{11}${School of Physics and Astronomy, University of Minnesota, Minneapolis, MN 55455, USA}
$^{12}${Department of Physics, California State University - East Bay, Hayward, CA 94542, USA}
$^{13}${School of Physics and Center for Relativistic Astrophysics, Georgia Institute of Technology, 837 State Street NW, Atlanta, GA 30332-0430}
$^{14}${School of Physics, National University of Ireland Galway, University Road, Galway, Ireland}
$^{15}${DESY, Platanenallee 6, 15738 Zeuthen, Germany}
$^{16}${Physics Department, McGill University, Montreal, QC H3A 2T8, Canada}
$^{17}${Santa Cruz Institute for Particle Physics and Department of Physics, University of California, Santa Cruz, CA 95064, USA}
$^{18}${Department of Physics and Astronomy, University of Utah, Salt Lake City, UT 84112, USA}
$^{19}${Department of Physics and Astronomy, University of Alabama, Tuscaloosa, AL 35487, USA}
$^{20}${Department of Physics and Astronomy, University of Iowa, Van Allen Hall, Iowa City, IA 52242, USA}
$^{21}${Institute of Particle and Cosmos Physics, Universidad Complutense de Madrid, 28040 Madrid, Spain}
$^{22}${Department of Physics and Astronomy, University of California, Los Angeles, CA 90095, USA}
$^{23}${Institute of Physics and Astronomy, University of Potsdam, 14476 Potsdam-Golm, Germany}
$^{24}${Department of Physical Sciences, Munster Technological University, Bishopstown, Cork, T12 P928, Ireland}
$^{25}${Department of Physics and Astronomy, Purdue University, West Lafayette, IN 47907, USA}

\end{document}